\documentclass[12pt]{article}
\usepackage{graphics}
\usepackage{setspace}

\input tcilatex
\QQQ{Language}{
American English
}

\begin{document}

\title{2D-Drop model applied to the calculation of step formation energies on a
(111) substrate.}
\author{M.I. Rojas, G. E. Amilibia, M.G. Del P\'{o}polo and E.P.M. Leiva\thanks{
Corresponding author. Fax 54-351-4334189; e-mail: eleiva@fcq.unc.edu.ar } \\
{INFIQC. Unidad de Matem\'{a}tica y F\'{\i}sica}\\
{Fac. Cs. Qu\'{\i}micas,} {Universidad Nacional de C\'{o}rdoba}\\
{Ciudad Universitaria, 5000 }\\
{C\'{o}rdoba, Argentina.}}
\maketitle

\begin{abstract}
A model is presented for obtaining the step formation energy for metallic
islands on (111) surfaces from Monte Carlo simulations. This model is
applied to homo (Cu/Cu(111), Ag/Ag(111)) and heteroepitaxy (Ag/Pt(111))
systems. The embedded atom method is used to represent the interaction
between the particles of the system, but any other type of potential could
be used as well. The formulation can also be employed to consider the case
of other single crystal surfaces, since the higher barriers for atom motion
on other surfaces are not a hindrance for the simulation scheme proposed.

\end{abstract}

\section{Introduction}

The formation of metal islands on a metal surface represents an important
stage in the growth of a phase in the case of homoepitaxy, or the appearance
of a new one in the case of heteroepitaxy. In any case, this topic has
interested whole generations of theoreticians and experimentalists, and is
currently the subject of intense research. Just trying to cite relevant work
in this field would require by itself a whole extensive review, so we just
mention briefly the work which may be connected directly to the present
letter.

Concerning homoatomic systems, recent experimental work has allowed the
accurate determination of the equilibrium shape of two dimensional metal
islands on different single crystal surfaces\cite{Giesen2001}. The analysis
by the inverse Wulff construction allowed the determination of the ratio of
the free energies per step length of the A- and B- type steps on Cu(111) and
Ag(111) surfaces, as well as the angle dependence of the step free energy
for Cu(100), Cu(111) and Ag(111). The analysis of the experimental data
using Ising models worked quite well for the (111) surfaces but were not
helpful in general for the (100) surfaces. On the other hand, an alternative
method for determining the kink energy from Arrhenius plots was developed,
which did not rely on a specific model for the interactions between the
atoms but on thermodynamic arguments. An important quantity also emerging
from these studies was the step energy per atom, which was obtained for
straight and 100 \% kinked steps, thus yielding a detailed picture of the
energetics of these systems.

In the case of heteroatomic systems, the interatomic interactions and the
adsorbate/substrate misfit determine the structure and the growth mechanism
of the monolayer and subsequent adlayers\cite{1}. Although the experimental
data on these systems is extensive, information concerning the step
formation energy is scarce. This quantity has been estimated from
electrochemical measurements by Xia et al \cite{Xia} for the system
Cu/Au(111), finding a value of 0.4-0.5 eV/(step atom). As we shall see
below, in the case of deposition of Ag on Pt(111) \cite{Röder_PRL} a rough
estimation of this quantity can also be made from the temperature at which
the fragmentation of islands occurs.

It is the purpose of the present work to formulate a model for the
estimation of step formation energies from Monte Carlo simulations, and
perform calculations for some typical systems. Our model includes the
natural relaxation of the system due to temperature effects, which coupled
to the misfit between substrate and adsorbate may produce some singular
effects in certain systems. Due to the simplified potential employed, we do
not expect to make accurate quantitative predictions, but we think the
present results will motivate further research in the field. The systems
studied were Cu/Cu(111), Ag/Ag(111) and Ag/Pt(111).

\section{The Drop model}

In the present calculations we focus on flat islands that are only one atom
thick. Both the energy of adsorption per atom in the bulk of the island and
edge may be of interest. In order to separate the two contributions, we have
adapted the drop model \cite{drop_model} that is often used for the
calculation of surface energies to the two-dimensional situation. This model
permits to obtain the binding energy of the atoms in the monolayer $u_s$ and
the binding energy of the atoms at the border of the island $u_b$ from the
adsorption energy per atom $u_t$, the number of border atoms $N_b$ and the
number of inner atoms $N_s$ that can be obtained for the Monte Carlo
simulation.

Let $U_T$ be the excess of energy of a system composed of an island adsorbed
on a substrate surface $M$:

\[
U_T=(U_{A/M}-U_M) 
\]
where $U_{A/M}$ is the energy of the adsorbed-substrate system and $U_M$ is
the energy of the substrate without adsorbate.

If we consider the adsorbed submonolayer like a 2-D drop, $U_T$ can be
expressed in terms of the edge atoms $N_b$ and the number of atoms which
make up the surface $N_s$ as:

\begin{equation}
U_T=u_sN_s+u_bN_b  \label{ut2}
\end{equation}

Let $N$ be the total number of atoms in the island:

\[
N=N_{s}+N_{b} 
\]

If we replace $N_s$ from this equation into equation \ref{ut2} , we obtain:

\[
U_T=u_sN+(u_b-u_s)N_b 
\]
And dividing by $N$ we get: 
\begin{equation}
u_t=\frac{U_T}N=u_s+(u_b-u_s)\frac{N_b}N  \label{ut3}
\end{equation}

Thus, a linear relationship is expected between the average values $<u_t>$
and $\frac{<N_b>}N$ for simulations employing islands of different sizes.
The $u_s$ and $u_b$ were obtained using least squares to calculate the
ordinate and slope of the straight line respectively.

The ordinate $u_s$, corresponding to the limit $\frac{N_b}N\rightarrow 0$
should approach the binding energy of a monolayer. The slope $(u_b-u_s)$ is
the border formation energy per atom. This should always be a positive
quantity and its amount indicates the energy change required to bring an
atom from the surface to the edge of the island.

.

\section{Model and Computations}

The substrate surface was assumed to be smooth without considering steps or
reconstruction. The substrate employed in Monte Carlo simulations was a
4-layer slab with 480 atoms per layer to represent (111) crystalline
surface. The atoms in the first and the second planes of the substrate
located within a circular area containing ca. 250 atoms were allowed to vary
the positions during the simulation. The remaining substrate atoms were
fixed to their bulk equilibrium configuration to emulate the presence of a
semi-infinite crystal. On this substrate, submonolayers of an initially
circular shape containing between 12-150 atoms were adsorbed to emulate
islands of different sizes.

The potentials employed were given by the embedded atom method (EAM) which
takes into account many body interactions characteristic of the metallic
systems \cite{6},\cite{7}. The total energy is expressed in terms of two
contributions:

\[
E_{tot}=\sum_{i=1}^NF_i(\rho _{h,i})+\frac 12\sum_i\sum_{j\neq i}\phi
_{ij}(r_{ij}) 
\]
where $\rho _{h,i}$ is the host electronic density at $i$ atom position, $%
F_i(\rho )$ is the embedding energy and $\phi _{ij}(r_{ij})$ is the
core-core pair repulsion between $i$ and $j$ atoms separated at a $r_{ij}$
distance. In all cases we used the parametrization of Foiles et al \cite{7}.

All the Monte Carlo simulations were performed in the (NVT) ensemble at
300K. We considered 5000 equilibration steps followed by 15000 production
steps.

In order to let the adatoms overcome high energy barriers, like those for
the displacement on the surface or the detachment from a cluster, we allowed
these adatoms to perform long jumps, with displacements $\Delta r $:

\[
\Delta r=n_1\overrightarrow{s_1}+n_2\overrightarrow{s_2} 
\]
where $n_1$ and $n_2$ are integers and $\overrightarrow{s_1}$ and $%
\overrightarrow{s_2}$ are the primitive vectors of a two-dimensional Bravais
lattice. These ``long'' jumps are important to get a proper equilibration of
the system. ``Short'' jumps, as usually employed in Monte Carlo simulations
where the positions of the particles are varied continuously, are also
present in our studies to describe the vibrational motion of the adsorbate
atoms in the neighborhood of their equilibrium positions. This simulation
method was employed in a previous work \cite{8} to study Pd submonolayer
growth on different monocrystalline surfaces, yielding results in reasonable
agreement with experimental data of literature.

\section{Results and Discussion}

The equilibrium shapes of two dimensional islands have been the subject of
extensive research in both the experimental and the theoretical fields. As
mentioned above, a recent work of Giesen et al has analyzed this problem in
detail\cite{Giesen2001}. While our current computer capabilities do not
allow the performance of simulations able to predict the shapes of island
with sizes in the order of tenths of nanometers, we shall tackle the problem
of predicting step formation energies.

We performed simulations for the homoatomic systems Cu/Cu(111) and
Ag/Ag(111) and for the heteroatomic Ag/Pt(111) system for a series of
islands of different sizes. The values of $<u_t>$ and $\frac{<Nb>}N$ were
obtained from the simulation production steps for the different submonolayer
sets and are plotted in Figure 1. We also show there the $<u_t>$ values
obtained from a simulation with a monolayer. We have plotted these monolayer
energies assigning them the value $\frac{<Nb>}N=0.$ For the larger $N$
employed in our simulations, we observe a linear behavior for the three
systems studied. A least square fitting in the range \textit{\ }$0.2<\frac{%
<Nb>}N<0.4$ using eqn. (\ref{ut3}) delivered the $u_s$ and $u_b$ reported in
Table 1. This set corresponds to the larger islands simulated. For
Cu/Cu(111) and Ag/Ag(111) the step formation energies $u_b-u_s$ are in the
order of 0.2-0.3 eV, which are close to the values found by Giesen et al. 
\cite{Giesen2001}. In the case of the Ag/Pt(111) system, an unusually low
step formation energy is found. This system presents an important positive
misfit $(4.3\%)$ which generates a strong compressive stress in the
pseudomorphic ($1\times 1$) monolayer\cite{Ibach}. R\"{o}der et al\cite
{Röder_PRL} have shown that when islands are grown at low temperatures and
subsequently annealed, a remarkable morphology transition is found around
300-350 K. At these temperatures, island fragmentation occurs via edge
roughening and kink evaporation, so that at 400 K all large Ag islands are
eventually disintegrated. Generally speaking, island disintegration should
be expected to take place when the step formation energy is of the order of
kT. Our estimation of step formation energy of ca. 0.085 eV, small compared
with those of other systems, delivers a disintegration temperature (990 $%
^{\circ }$K) which results far above the experimental value. While this
disagreement could be roughly attributed to the approximate nature of the
potential employed, close observation of figure 1 shows that the linear
extrapolation of the $<u_t>$ vs $\frac{<Nb>}N$ curve to $\frac{<Nb>}%
N\rightarrow 0$ shows the strongest deviation with respect to the $<u_t>$ of
the monolayer for the Ag/Pt(111) system. Thus, the question arises if some
of the approximations involved in the present model are responding bad for
this system.\ In this sense, the extrapolation to $\frac{<Nb>}N\rightarrow 0$
to obtain the ordinate $u_s$ from equation (\ref{ut3}) implies the
assumption that the island is roughly composed of two sorts of atoms, say
''border '' and ''inner '' atoms. In order to learn how fast the atoms
inside the island approach the behavior of atoms in a monolayer ( ''inner ''
atoms), we have plotted in figure 2 the distance between nearest neighbors
at the center of the monolayer as a function of the total number of atoms in
the island. In the case of the system Ag/Ag(111), we see that for relatively
small islands the atoms at the center already reflect the behavior of the
monolayer. In the case of the Cu/Cu(111) system, the approach to the
monolayer behavior is more sluggish, but is practically reached for the
large island in our simulation. On the other hand, we see that in the case
of the Ag/Pt(111) system the behavior of the inner atoms is far from that of
Ag atoms in a (1x1) monolayer on Pt(111). These results indicate that small
Ag islands on Pt(111) are considerably expanded with respect to the (1x1)
monolayer, becoming more and more compressed with growing island size,
probably reaching finally a (1x1) structure with the substrate for sizes
that are beyond our simulation conditions. In this way, the binding energy
of the atoms in the monolayer is not accurately estimated from the
extrapolation in figure 1. On the other hand, the $u_s$values can be
properly obtained from a simulation employing a (1x1)\ adsorbed monolayer,
and the linear fitting of the points in figure 1 can be made by involving $%
u_b$ as unique free parameter. Table 2 shows $u_s$ values stemming from
monolayer simulations, as well as the $u_b$ obtained from linear fittings
taking $u_s$ from these monolayer simulations. The corresponding step
formation energies $u_b-u_s$ are also reported there. Comparison with Table
1 indicate that although the extrapolated $u_s$values were reasonable,
important changes occur in the step formation energy, especially in the case
of Ag/Pt(111), where $u_b-u_s$ has strongly decreased.

A last point that we have checked is the influence of relaxation in the
calculated adsorption and step formation energies. In Table 3 we show the
quantities obtained from static calculations, where the island and surface
atoms are not allowed to relax. As in the previous case, the most drastic
differences are found in the case of the step formation energy of the system
Ag/Pt(111).

For all the island sizes of Ag/Pt(111) considered in this work, we found an
outwards relaxation of the edge atoms of the islands, which confirms the
statement of Bromann et al \cite{3}concerning the fact that compressive
stress can be relieved by this way. However, we did not find pseudomorphic
growth in our islands, but rather expanded structures which become
progressively compressed as the islands increase their size. Simulations
using periodic boundary conditions showed that the (1$\times $1) structure
was stable under our simulation conditions. Thus, the picture that emerges
from the present calculations is that the increased size of the islands
leads to progressively compressed structures of Ag on Pt(111) until they
reach the (1$\times $1) structure of the monolayer. The actual behavior of
the experimental system is richer, with the appearance of partial
dislocations, yielding domain walls between fcc and hcp stacking \cite{3}.
In this respect, simulations with a much larger number of atoms are
desirable in order to investigate if the EAM potential is able to reproduce
such subtle features.

In contrast to the Ag/Pt(111) system, the other systems present inwards
relaxation at the edge, following the natural expectation from bond-order
conservation analysis, atoms at the edge are less coordinated, and they
increase the strength of the binding with each other.

We can summarize the present results by stating that the model we have
presented permits to obtain the step formation energy for metallic islands
on (111) surfaces in a straightforward way for homo and heteroepitaxy
systems from Monte Carlo simulations. We have used the embedded atom method
to represent the interaction between the particles of the system, but any
other type of potential could be used as well. The model can also be used
for other single crystal surfaces, since the higher barriers for atom motion
on other surfaces are not a hindrance for the simulation scheme proposed.

\section*{Acknowledgments}

This work was supported by PIP and PEI 86/98 CONICET, Program BID 802/OC-AR
PICT 06-04505, CONICOR, Secyt UNC, Argentina. We also wish to thank M.I.
Baskes for providing the tables with the potential functions. Language
assistance from P. Falcon is also acknowledged.\newpage\

\section{Tables}

\begin{center}
\begin{tabular}{|c|c|c|c|}
\hline
System & $u_s$ {[}eV{]} & $u_b$ {[}eV{]} & $u_b-u_s$ {[}eV{]} \\ \hline\hline
Cu/Cu(111) & -3.509$\pm $0.001 & -3.194$\pm $0.004 & 0.315$\pm $0.003 \\ 
\hline
Ag/Ag(111) & -2.817$\pm $0.001 & -2.583$\pm $0.004 & 0.234$\pm $0.003 \\ 
\hline
Ag/Pt(111) & -3.167$\pm $0.002 & -2.997$\pm $0.008 & 0.085$\pm $0.006 \\ 
\hline
\end{tabular}
\end{center}

Table 1: Binding energy per atoms in the monolayer $u_s$, binding energy per
atom at the edge of the island $u_b$ and step formation energy per atom $%
u_b-u_s$ for different systems. These values were obtained from linear
fitting of the points in figure 1 in the range $0.2<\frac{<Nb>}N<0.4.\,$The
parameters fitted were $u_s$ and $u_b-u_s$ according to equation (\ref{ut3}%
). The Monte Carlo/EAM simulations were performed at 300 K.\\

\begin{center}
\begin{tabular}{|c|c|c|c|}
\hline
System & $u_s$ {[}eV{]} & $u_b$ {[}eV{]} & $u_b-u_s$ {[}eV{]} \\ \hline\hline
Cu/Cu(111) & -3.498 & -3.216$\pm $0.002 & 0.283$\pm $0.002 \\ \hline
Ag/Ag(111) & -2.808 & -2.600$\pm $0.001 & 0.208$\pm $0.001 \\ \hline
Ag/Pt(111) & -3.150 & -3.118$\pm $0.002 & 0.032$\pm $0.002 \\ \hline
\end{tabular}
\end{center}

Table 2: Binding energy per atom in the monolayer $u_s$, binding energy per
atom at the edge of the island $u_b$and step formation energy per atom $%
u_b-u_s$ for different systems. The $u_s$ were obtained from monolayer
simulations. The $u_b$ values were obtained from linear fitting of the
points in figure 1 in the range $0.2<\frac{<Nb>}N<0.4$ according to equation
(\ref{ut3}). The Monte Carlo/EAM simulations were performed at 300 K.\\.%
\newpage 

\begin{center}
\begin{tabular}{|c|c|c|c|}
\hline
System & $u_s$ {[}eV{]} & $u_b$ {[}eV{]} & $u_b-u_s$ {[}eV{]} \\ \hline\hline
Cu/Cu(111) & -3.555$\pm $0.003 & -3.21$\pm $0.01 & 0.350$\pm $0.007 \\ \hline
Ag/Ag(111) & -2.861$\pm $0.002 & -2.597$\pm $0.008 & 0.264$\pm $0.006 \\ 
\hline
Ag/Pt(111) & -3.164$\pm $0.002 & -3.001$\pm $0.003 & 0.163$\pm $0.001 \\ 
\hline
\end{tabular}
\end{center}

Table 3: Binding energy per atom in the monolayer $u_s$, binding energy per
atom at the edge of the island $u_{b\text{ }}$and step formation energy per
atom $u_b-u_s$ for different systems, as obtained from static EAM
calculations. These values arise from a linear fitting of the adsorption
energy per atom in the range $0.2<\frac{<Nb>}N<0.4.\,$The parameters fitted
were $u_s$ and $u_b-u_s$ according to equation (\ref{ut3}).\newpage\

\begin{figure}[t]
{\centering \resizebox*{12cm}{12cm}{\rotatebox{-90}{\includegraphics{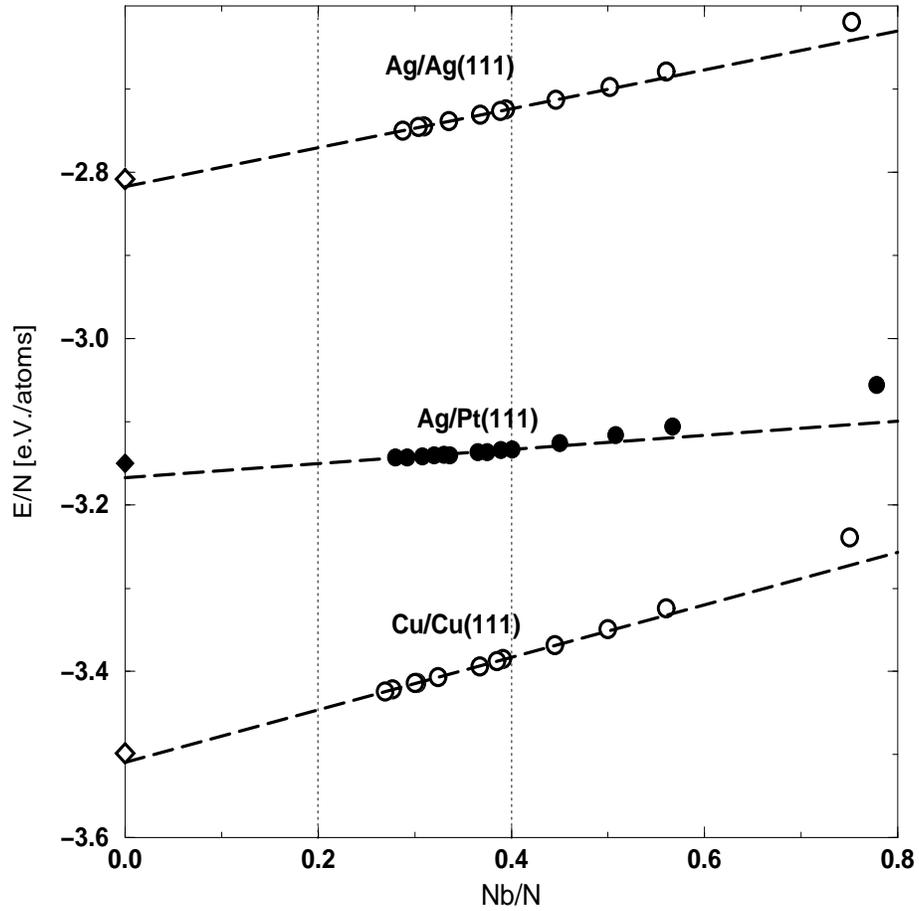}}} \par} 
\caption{Adsorption energy per atom $<u_t>$ vs fraction of border atoms $%
\frac{<Nb>}N$curve for different adsorbate(island)/substrate(111) systems.
The points at $\frac{<Nb>}N=0$ correspond to monolayer simulations.}
\end{figure}

\begin{figure}[t]
{\centering \resizebox*{12cm}{12cm}{\rotatebox{-90}{\includegraphics{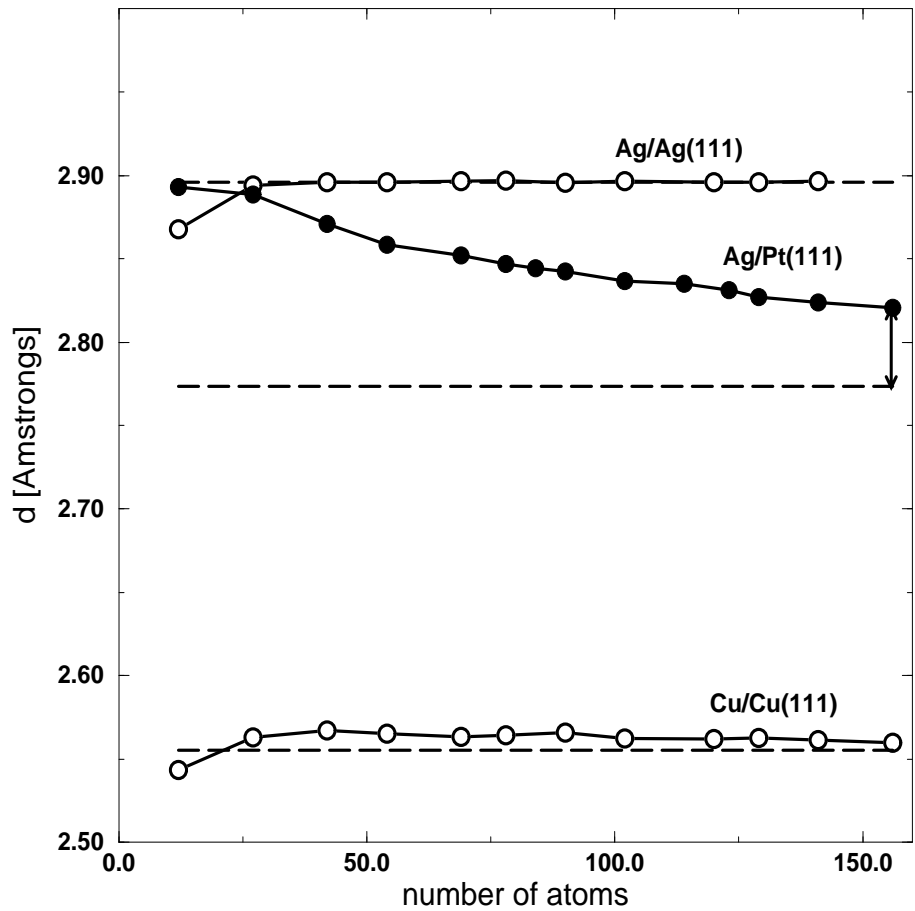}}} \par}

\caption{Nearest neighbor distance at the center of the island as a
function of the total number of island atoms for the systems considered in
this work.}
\end{figure}

\end{document}